# An Efficient Paging Algorithm for Multi-Carrier CDMA System

Sheikh Shanawaz Mostafa[1], Khondker Jahid Reza[2], Gazi Maniur Rashid[3], Muhammad Moinuddin[4], Md. Ziaul Amin[5] and Abdullah Al Nahid[6]

[1,2,3,4,5,6] Electronics and Communication Engineering Discipline, Khulna University,
Khulna-9208, Bangladesh.

**Abstract**
To cope with the increasing demand of wireless communication services multi-carrier systems are being used. Radio resources are very limited and efficient usages of these resources are inevitable to get optimum performance of the system. Paging channel is a low-bandwidth channel and one of the most important channels on which system performance depends significantly. Therefore it is vulnerable to even moderate overloads. In this paper, an efficient paging algorithm, Concurrent Search, is proposed for efficient use of paging channel in Multi- carrier CDMA system instead of existing sequential searching algorithm. It is shown by the simulation that the paging performance in proposed algorithm is far better than the existing system.

**Keywords:** *Absorbing Markov Chain, Concurrent Search, Erlang C Formula, Multi-Carrier, Paging.*

## 1. Introduction

In order to deliver services to a Mobile Station (MS) the cellular network must be able to track MS in the network. The forward-link communication channels are used to page and transmit system overhead messages to the MS is called paging channel [1].

Efficient utilization of limited resources is required to support the growing demands of personal communication services. To utilize the paging channel efficiently, many researchers have been studying different techniques. One of the concepts is dividing a location area into paging zones is described in [8] and [9]. Another research has proposed to increase paging efficiency-MS will be paged on the last registered cell and subsequently to other cells in the location area if necessary [4]. A mobility tracking scheme that combines a movement-based location update policy with a selective paging scheme is proposed in [5].

In CDMA system, an MS accesses the system on a particular carrier frequency [10]. A single carrier frequency occupies 1.25MHz bandwidth contains seven paging channels. When a carrier reaches its capacity limit, it is desirable to increase the system capacity. One way of achieving this is to increase the number of carriers to greater than one. When more than one carrier is used then, the system is referred to as multi-carrier system.

In existing multi carrier system, an MS is capable of tuning to only one of the carrier frequencies at any instant of time. Because of this, the MS is only listening to one of the carrier frequency paging channels at a time. In order to page an MS, the paging message must be sent on the paging channels of all of the carrier frequencies to ensure that it is sent on the paging channel the MS is listening to. To do this the paging message must be duplicated for each of the paging channels on which the message is to be sent [10]. The existing system can't search more than seven users simultaneously. This results in an inefficient utilization of paging channel recourses.

In this paper, a concurrent search algorithm is proposed which can be used to search more than seven users simultaneously in multiple carriers. The concurrent search algorithm has been used for paging in cells [6]. But the novelty of our paper is the adoption of concurrent search algorithm in multiple carrier CDMA system rather then cells. The system performance has been analyzed by using absorbing Markov chain and Erlang C formula.

## 2. Sequential Search Algorithm

This paper focused concurrent search algorithm based on the problem of sequentially locating a number of users. Suppose there is n number of carriers in the network and k mobile users to be located. In existing multicarrier CDMA system a straight forward sequential paging scheme has been used to locate user i among k number of users. We assume that the user must exists in the network so $\sum_{j=1}^{n} P(i,j) = 1$ $\forall$ i ∈ {1,2, ...k}. When 1 ≤ c ≤ n where P(i,j) is the probability of existence of user i in $j^{th}$ carrier. Whereas F (i,c) is the indicator of finding user and that is F(i,c) ∈ {0,1} if the $i^{th}$ mobile user has located at the $c^{th}$ carrier then F(i,c) = 1 elsewhere F(i,c) = 0. To clarify the existing problem we use simplify algorithm of sequential search.

**Step 0:** Initially F(i,c) = 0 where 1 ≤ i ≤ k and 1 ≤ c ≤ n
**Step 1:** [flood the message]
for all 1 ≤ c ≤ n
Copy and send the paging message for user i
**Step:3**
If F(i,c) = = 1
exit

This technique requires k*n paging channels by using concurrent paging scheme it is possible to reduce average use of paging channels.

Consider an example to demonstrate the effectiveness of concurrent search scheme. Suppose that there are only two carriers in the system with same number of users. Among them two users: X and Y are to be found. Assume that the system has no previous knowledge about the location of the users: each user can exist in any of the two carriers with probability 0.5. Using sequential paging, the system would page and locate user X to the both carriers in the first attempt. Then, the system would page and locate user Y to the both carriers in the second attempt. Thus total of (2+2) four pages would be required for finding two users.

On the other hand, if the system is to page in the first attempt user X in carrier 1 and for user Y in carrier 2 simultaneously, then there is 50% probability that any of the two users would be found in the first attempt. If a user Y is not found at first attempt, the system would page for the user in the "other" carrier in the second attempt. Thus, (2+1) 3 paging messages are needed for the two users. On the average, 1.5 pages would be required per user, a saving of 25% over the sequential paging.

## 3. Concurrent Search Algorithm

The proposed concurrently searching algorithm for locating $i^{th}$ user in n carriers is described here. To minimize the paging cost, a methodology is described in [7], it is stated that the higher probability of finding users must be paged before the lower probability. This is also true for proposed carrier concept. From Fig. 1 and Fig. 2, it is visualized that if the priority is high then the cost (time) is low.

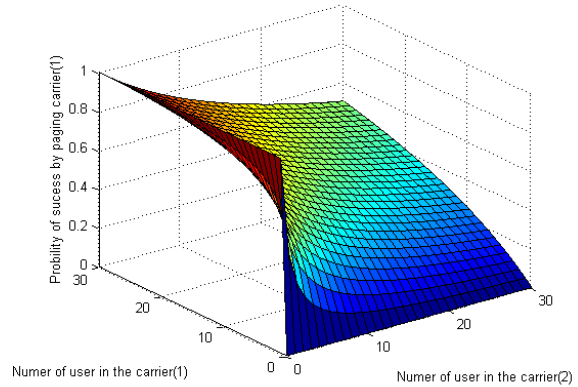

Fig. 1 Probability of paging success

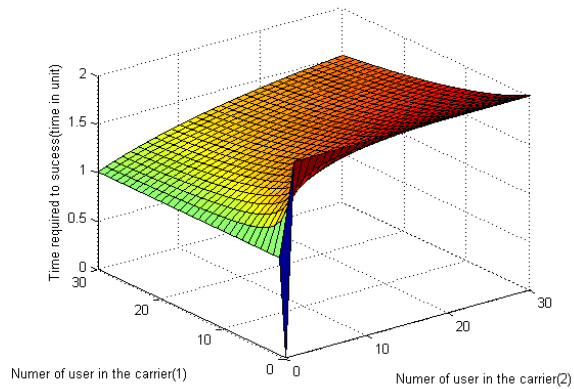

Fig. 2 Cost (time) of paging

It is assumed that the mobile users do not move to other carriers during searching. Terms used in probability algorithm are P (i,j) be the probability that a mobile user i is in carrier j. When a call to mobile user i arrives, generally assumed that P(i,j) > 0 $\forall$ i ∈ {1,2,3.....k}, j ∈ {1,2,....n}. If the total number of users in $j^{th}$ carrier is $U_j$ then probability $P(i, j) = \dfrac{U_j}{\sum_{c=1}^{n} U_c}$, and $\sum_{j=1}^{n} P(i,j) = 1$ $\forall$ i ∈ {1,2, ...k}. When 1 ≤ c ≤ n S(i,c) ∈ {0,1}, if the $i^{th}$ mobile user has not located at the $c^{th}$ carrier then S(i,c) = 0 elsewhere S(i,c) = 1, where 1 ≤ i ≤ k. And the matrix $B_i[c]$ stores up to date priority information about carrier for paging. Which can be found from the probabilistic information P(i,j) by using any sorting algorithm.

The proposed concurrent search algorithm is presented as follows.
**Step 0:** Initially, S(i,c) = 0, for all $1 \leq i \leq k$
**Step 1:** [Initializing carrier]
  set c = 1
**Step 2:** [Carrier Choice]
  (a) If $B_i[c] \neq 0$ and $B_i[c]$ is free then perform step 3 to page mobile user i in the carrier c.
  (b) else if c+1 ≤ n
    set c = c+1 and go to step 2
  (c) else if c+1 > n and
    then go to step1
  (d) else go to step 4
**Step 3:** [Page]
  (a) if S(i ,c) = = 1
    go to step 4
  (b) else set $B_i[c]$ = S(i,c), c = c+1 and go to step 2
**Step 4:** exit

Here are some remarks related to above algorithm.
Step 0, the initial states, S(i,c) = 0,which represents no MS has been located yet.

In step 1, the 1st carrier is chosen among the 'n' number of the carriers from priority array of $B_i[c]$.

In step 2(a), paging message will be sent (actually goes to step 3) if and only if c is not paged yet and c is free. Here it should be noted that, if the user is not found in that particular carrier is assigned that carrier $B_i[c]$ as 0 to obstruct multiple page on the same carrier.

In step 2(b), if step 2(a) is not satisfied then next carrier number will be checked if it is less than or equal total number of carrier. Again step 2 will be performed.

In step 2(c), if the highest number of carrier is exceeded and all the carriers are paged then the system goes to step 2(d). Otherwise step 1 will be performed.

Step 2(d) will be performed when 2(a) or 2(b) or 2(c) steps are not satisfied.

In step 3, paging procedure is mainly processed.

In step 3(a), if the desired user is found then S(i,c) is set to 1.

In step 3(b), if the desired user is not found then S(i,c) is copied to Bi[c]and carrier is increased by 1. System goes to step 2 again.

In step 4, exit.

## 4. Performance Evaluation

For analyzing the performance, consider there are two carriers in the system. So that, for CDMA system there are 14 (2*7) paging channels. Let, carrier 1 has M number of users and carrier 2 has N number of users so the total number of users in the system is M+N. Probability of finding a particular user in carrier 1 is $\frac{M}{M+N}$ and in carrier 2 is $\frac{N}{M+N}$. If the system first paged at carrier 1 for the desired user then the state diagram will be like Fig. 3

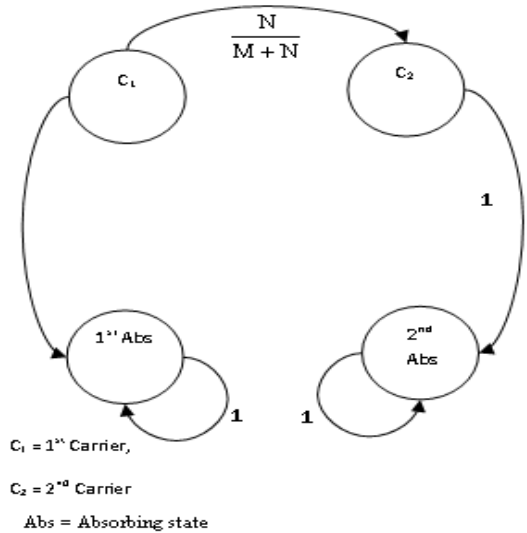

Fig. 3 State Diagram

In Fig. 3, if the user is succeeded in the 1st carrier, then goes to 1st Abs state with the probability $\frac{M}{M+N}$. But if the user is not succeeded at 1st Carrier then it goes to 2nd Carrier with a probability $\frac{N}{M+N}$. As it is assumed earlier that the desired user must be present in the system, so user must be succeeded in the 2nd Carrier. Then the probability of going from 2nd Carrier to 2nd Absorbing state is 1. Here, 1st Abs and 2nd Abs are the absorbing state which mean that the paging request no longer in the system. A user request goes to outside of the system through these states. So the canonical form of the state diagram is.

$$P = \begin{array}{c} C_1 \\ C_2 \\ 1^{st} Abs \\ 2^{nd} Abs \end{array} \begin{bmatrix} \overset{C_1}{0} & \overset{C_2}{\dfrac{N}{M+N}} & \overset{1^{st} Abs}{\dfrac{M}{M+N}} & \overset{2^{nd} Abs}{0} \\ 0 & 0 & 0 & 1 \\ 0 & 0 & 1 & 0 \\ 0 & 0 & 0 & 1 \end{bmatrix}$$

It is assumed that the user spends a unit time in single carrier. The average paging (service) time that the user stays in the service can be calculated by using absorbing Markov chains [2]. That is $1+\dfrac{N}{M+N}$ or $2-\dfrac{M}{M+N}$ from which it is realized that M should be always higher than N (shown in Fig 2.) for reducing the cost (time). There are seven paging channels of each carrier. So that, greater than seven users can't access the 1$^{st}$ carrier simultaneously as a result they go to 2$^{nd}$ carrier. And their average time will be $1+\dfrac{N}{M+N}$ or $2-\dfrac{M}{M+N}$. The average time a user spends in the service can be found by calculating the expected value of user time which is 1.5 units. The system can be imagined with 14 parallel servers having queue [12]. Then its probability that all servers are busy can be found from Erlang C formula [3], [11] is,

$$\text{Pr[C channels are busy]} = \dfrac{A^c}{A^c + C!(1-\dfrac{A}{C})\sum_{k=0}^{C-1}\dfrac{A^k}{k!}} \quad (1)$$

Where A is total offered traffic which is $\lambda/\mu$, where λ is incoming traffic rate and μ is the service rate. C is the number of parallel servers comparing with proposed system it is channels. So the probability that all channels are busy in proposed system for 14 channels is $P_o$

$$P_0 = \dfrac{A^{14}}{A^{14} + 14!(1-\dfrac{A}{14})\sum_{k=0}^{13}\dfrac{A^k}{k!}} \quad (2)$$

Average Wait for All users

$$AWA = \dfrac{P_0}{\mu \times (14-A)} \quad (3)$$

Average Wait for Delayed users

$$AWD = \dfrac{1}{\mu \times (14-A)} \quad (4)$$

Average Time in the System

$$T = AWA + \dfrac{1}{\mu} \quad (5)$$

## 5. Simulation

For two carriers service rate of sequential search is $\mu_s = \mu$ where the service rate of concurrent search is $\mu_c = 1.5\mu$. Here $\mu$ is the service rate of single carrier. The is the blocking probability can be compared by Eq. (2) (Fig 4) and other performance parameter like average wait time for all users, average wait time for delayed users, average service time in the system can be analyzed by Eq. (3) (Fig 6), Eq. (4) (Fig 5) and Eq.(5) (Fig 6) sequentially.

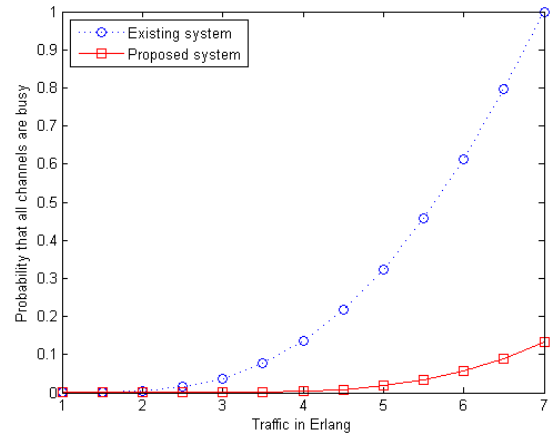

Fig. 4 Probability Comparison of busy channels for two carriers.

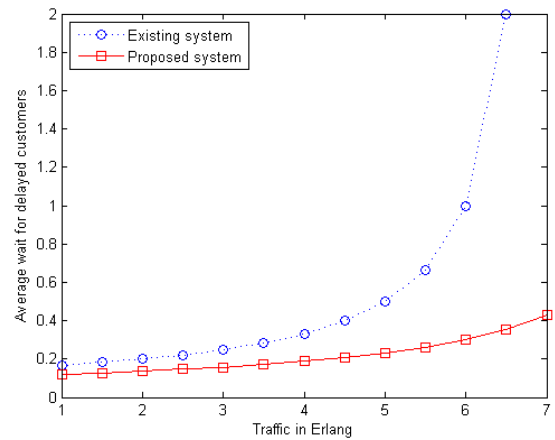

Fig. 5 Comparison of average wait for delayed users between proposed and existing system for two carriers.

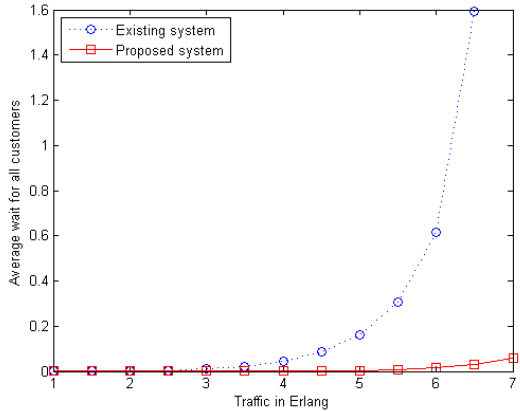

Fig. 6 Comparison of average waits for all customers between proposed and existing system for two carriers

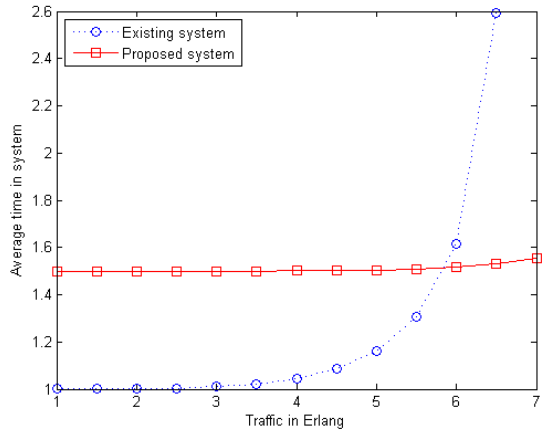

Fig. 7 Comparison of total average time between proposed and existing system for two carriers.

## 6. Conclusions

The proposed concurrent search algorithm for paging mobile users in CDMA system based on the probabilistic information about the carriers. The reduction of using paging channel due to such a concurrent search can be quite extensive; shows on simulation results. In the proposed system, traffic handling capability is higher with a little bit sacrifice in servicing time. But this drawback is also abolished for higher traffic. So the proposed system will work better in the peak hour with a large incoming traffic with same resources.

### References.

[1] Saleh Faruque, Cellular Mobile System Engineering, Artech House Publishers, New ed., 2007-2008.
[2] Charles M. Grinstead, J. Laurie Snell, Introduction to Probability, 2nd ed., American Mathematical Society, 1991.
[3] Theodre S. Rappaport, "Wireless Communications", 2nd ed., Prentice-Hall of India, 2002.
[4] G. L. Lyberpoulos, J.G. Markoulidakis, D.V. Polymeros, D.F. Tsirkas, and E.D. Sykas, "Intelligent paging strategiesfor 3rd generation mobile telecommunication systems," IEEE/ACM Trans. Veh. Technol., vol 44, pp. 573-553, Aug. 1995.
[5] I.F. Akyildiz, J.S. M. Ho, and Y.-B. Lin, "Movement-based locationupdate and selective paging for PCS networks," IEEE /ACM Trans. Networking, vol. 4, pp. 629-638, Aug. 1996.
[6] Rung-Hung Gue and Zygmunt J. Haas "Concurrent Search of Mobile Users in Cellular Networks", IEEE/ACM Transactions on Networking, vol. 12, no. 1, Feb 2004
[7] C. rose and R. Yates, "Minimizing the average cost of paging under delay constraints," ACM /Kluwer Wireless Networks, vol. 1, no.2, pp. 211-219, 1995.
[8] D. Munoz-Podrguez, "Cluster paging for traveling subscribers," in IEEE Vehicular Technology Conf. 1990, May, 6-9, pp. 748-753.
[9] D. Plassmann, "Location management strategies for mobile cellular networks of 3rd generation," IEEE Vehicular Technology Conf., June 1994, pp. 649-653.
[10] William E. Illidge, "CDMA Multiple Carrier Paging Channel Optimization", Patent number: US 6,542,752; Date of Patent Apr. 1, 2003.
[11] Erlang C queuing model (2010) Erlang Library for Excel. [Online].
Available:http://abstractmicro.com/erlang/helppages/mod-c.htm
[12] CDMA2000 1X paging Optimization Guide, Huawei Technologies Co., Ltd., 2004.

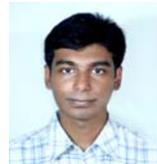

**Sheikh Shanawaz Mostafa** completed B.Sc. Engineering in Electronics and Communication, from Khulna University-9208, Khulna, Bangladesh. His current research interests are: Wireless communication, Modulation techniques and Biomedical signal processing. He has seven papers, published in different local and international recognized journal and proceedings of conference.

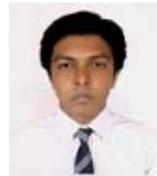

**Khondker Jahid Reza** is serving as an Engineer at Advance Data Networks System Ltd. He has completed his B.Sc. in Electronics and Communication Engineering Discipline in Khulna University, Khulna, Bangladesh. His current research interest is wireless communication, modulation and sensor networks. He has two papers, published in international recognized journal.

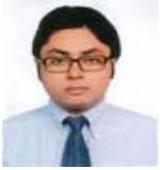

**Gazi Maniur Rashid** is serving as an RF Engineer at Metro global Telecom Service Ltd. He has completed his B.Sc. in Electronics and Communication Engineering Discipline in Khulna University, Khulna, Bangladesh. His current research interest is wireless communication, modulation techniques, channel coding and fading. Previously his paper published in an international recognized journal.

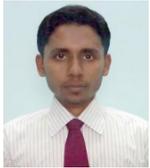

**Muhammad Moinuddin** is serving as an Assistant Network Engineer at Dhakacom Ltd. He has completed his B.Sc. in Electronics and Communication Engineering Discipline in Khulna University, Khulna, Bangladesh. His current research interests are: wireless communication and various modulation techniques.

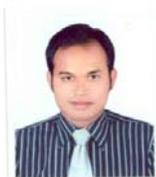

**Md. Ziaul Amin** is currently working as an Assistant Professor at the Khulna University, Khulna, Bangladesh. He obtained his B.Sc. in Electronics and Communication Engineering from same university. Previously, he worked as a System Engineer, planning, at RanksTel Bangladesh Ltd. since 10.11.07 to 08.09.08. His current research interests are: Digital Signal Processing and Radio Network Planning. He has three papers, published in international recognized journal.

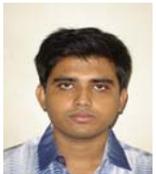

**Abdullah Al Nahid** is now working as a faculty member of Electronics and Communication Engineering (ECE) discipline at Khulna University, Khulna, Bangladesh. He completed his B.Sc. in ECE from the same university. He is working in the field of Digital Signal Processing, Robotics and Communication. He has more than ten papers, published in different local and international recognized journal and proceedings of conference.